\documentclass[sigconf]{acmart}
\AtBeginDocument{%
  \providecommand\BibTeX{{%
    \normalfont B\kern-0.5em{\scshape i\kern-0.25em b}\kern-0.8em\TeX}}}

\copyrightyear{2021}
\acmYear{2021}
\setcopyright{acmcopyright}\acmConference[Web3D '21]{The 26th International Conference on 3D Web Technology}{November 8--12, 2021}{Pisa, Italy}
\acmBooktitle{The 26th International Conference on 3D Web Technology (Web3D '21), November 8--12, 2021, Pisa, Italy}
\acmPrice{15.00}
\acmDOI{10.1145/3485444.3487644}
\acmISBN{978-1-4503-9095-8/21/11}

\usepackage{color}
\definecolor{lightgray}{rgb}{0.95, 0.95, 0.95}
\definecolor{darkgray}{rgb}{0.4, 0.4, 0.4}
\definecolor{editorGray}{rgb}{0.95, 0.95, 0.95}
\definecolor{editorOcher}{rgb}{1, 0.5, 0} % #FF7F00 -> rgb(239, 169, 0)
\definecolor{editorGreen}{rgb}{0, 0.5, 0} % #007C00 -> rgb(0, 124, 0)
\definecolor{orange}{rgb}{1,0.45,0.13}		
\definecolor{olive}{rgb}{0.17,0.59,0.20}
\definecolor{brown}{rgb}{0.69,0.31,0.31}
\definecolor{purple}{rgb}{0.38,0.18,0.81}
\definecolor{lightblue}{rgb}{0.1,0.57,0.7}
\definecolor{lightred}{rgb}{1,0.4,0.5}
\usepackage{listings}

\graphicspath{{figs/}} % where to search for the images

\lstdefinelanguage{JavaScript}{
  keywords={typeof, new, true, false, catch, function, return, null, catch, switch, var, const, let, async, await, if, in, while, do, else, case, break, from, default},
  ndkeywords={class, export, boolean, throw, implements, import, this, Promise},
  sensitive=false,
  comment=[l]{//},
  morecomment=[s]{/*}{*/},
  morestring=[b]',
  morestring=[b]"
}

\begin{document}

\title[{Resurrect3D: A Platform for Visualizing and Analyzing Cultural Heritage Artifacts}]{Resurrect3D: An Open and Customizable Platform for Visualizing and Analyzing Cultural Heritage Artifacts}

\author{Joshua Romphf}
\affiliation{%
  \institution{Digital Scholarship River Campus Libraries}
  \country{University of Rochester, USA}
}
\email{jromphf@library.rochester.edu}

\author{Elias Neuman-Donihue}
\affiliation{%
  \institution{Department of Computer Science, Department of History}
  \country{University of Rochester, USA}
}
\email{eneumand@u.rochester.edu}

\author{Gregory Heyworth}
\affiliation{%
  \institution{Department of English}
  \country{University of Rochester, USA}
}
\email{gregory.heyworth@rochester.edu}

\author{Yuhao Zhu}
\affiliation{%
  \institution{Department of Computer Science}
  \country{University of Rochester, USA}
}
\email{yzhu@rochester.edu}

\renewcommand{\shortauthors}{Romphf, et al.}

\begin{abstract}
Art and culture, at their best, lie in the act of discovery and exploration. This paper describes Resurrect3D, an open visualization platform for both casual users and domain experts to explore cultural artifacts. To that end, Resurrect3D takes two steps. First, it provides an interactive cultural heritage toolbox, providing not only commonly used tools in cultural heritage such as relighting and material editing, but also the ability for users to create an interactive ``story'': a saved session with annotations and visualizations others can later replay. Second, Resurrect3D exposes a set of programming interfaces to \textit{extend the toolbox}. Domain experts can develop custom tools that perform artifact-specific visualization and analysis.
\end{abstract}



%\keywords{datasets, neural networks, gaze detection, text tagging}


\maketitle

%!TEX root=paper.tex

\newcommand{\website}[1]{{\tt #1}}
\newcommand{\program}[1]{{\tt #1}}
\newcommand{\benchmark}[1]{{\it #1}}
\newcommand{\fixme}[1]{{\textcolor{red}{\textit{#1}}}}

\newcommand*\circled[2]{\tikz[baseline=(char.base)]{
            \node[shape=circle,fill=black,inner sep=1pt] (char) {\textcolor{#1}{{\footnotesize #2}}};}}

\ifx\figurename\undefined \def\figurename{Figure}\fi
\renewcommand{\figurename}{Fig.}
\newcommand{\figline}{{\vspace*{.05in}\hline}}

\newcommand{\Sect}[1]{Section~\ref{#1}}
\newcommand{\Fig}[1]{Figure~\ref{#1}}
\newcommand{\Tbl}[1]{Tbl.~\ref{#1}}
\newcommand{\Equ}[1]{Equ.~\ref{#1}}
\newcommand{\Apx}[1]{Apdx.~\ref{#1}}
\newcommand{\Alg}[1]{Algo.~\ref{#1}}

\newcommand{\specialcell}[2][c]{\begin{tabular}[#1]{@{}c@{}}#2\end{tabular}}
\newcommand{\note}[1]{\textcolor{red}{#1}}

\newcommand{\proj}{\textsc{Mesorasi}\xspace}
\newcommand{\mode}[1]{\underline{\textsc{#1}}\xspace}
\newcommand{\sys}[1]{\underline{\textsc{#1}}}

\newcommand{\no}[1]{#1}
\renewcommand{\no}[1]{}
\newcommand{\RNum}[1]{\uppercase\expandafter{\romannumeral #1\relax}}

\def\cA{{\mathcal{A}}}
\def\cF{{\mathcal{F}}}
\def\cN{{\mathcal{N}}}

% checkmark and xmark in the pifont package
%\newcommand{\cmark}{\ding{51}}
%\newcommand{\xmark}{\ding{55}}

\section{Introduction}

Art and culture, at their best, lie in the act of discovery: seeing what is hidden and rejecting the fallacy that what we see is all there is. Almost every great piece of art, artifact, or site has something hidden within it, invisible to the naked eye. Paintings have pentimenti --- rough drafts that an artist regrets and paints over; manuscripts have palimpsests or scratch-outs --- text that is covered up or overwritten; and archeological sites are a world of unexpected labyrinths.
%objects have difficult-to-see features that reveal how they are made, by whom and when; 

The discovery aspect of culture and art is inadequately addressed by museums today. Even with a guided tour, the museum experience is centered around receiving information rather than discovering new information. 3D digitization and visualization, however, are transforming this landscape, allowing user-centric exploration in both casual consumption and scientific research of cultural heritage.

\subsection{Design Principles}

In this paper, we describe Resurrect3D, an open platform for visualizing, analyzing, and ultimately freely exploring difficult-to-study artifacts. Resurrect3D is built around three key design principles.

\paragraph{Targeting Cultural Heritage.} While using a generic system architecture and supporting general 3D visualization, Resurrect3D is designed specifically with cultural heritage in mind. For instance, we support processing data from a range of data acquisition pipelines such as LiDAR scanning, multispectral imaging, and photogrammetry; similarly, we provide well-optimized tools to annotate, analyze, and visualize 3D objects. These features allow domain experts to not only interact with the artifacts but also ``see the invisibles'' and analyze artifacts using advanced algorithms.

%, ideally with low engineering cost and the porting overhead for every update of the application.

%, ranging from casual users that mainly use desktop PCs and smartphones to classroom and lecturing scenarios that make heavy use of emerging technologies such as AR and VR. 

\paragraph{Portability.} Our experience of developing visualization platforms for cultural heritage tells us that the end users are extremely diverse. Resurrect3D must be available across different computing platforms ranging from desktop PCs to smartphones and Virtual Reality devices. To this end, we made a judicious design decision to base the entire system on the Web technology stack (e.g., NodeJS, WebGL, WebXR), which is naturally platform-independent. The Web application will automatically get updated whenever the user launches the application, i.e., requesting the application through a URL, allowing for automatic security/privacy improvements.

%In contrast, many existing cultural heritage visualization platforms are developed for specific platforms, introducing significant friction on user experience when switching platforms and engineering overhead for the developers.

%To that end, Resurrect3D is built completely on top of the Web stack as a Web application, which is naturally platform-independent through its ``write-once, run-anywhere'' characteristic. 

%These custom tools should also have interoperability between them. For instance, a user may wish to add metadata to an object using the annotation tool, while also persisting settings from the custom shaders alongside that particular annotation. Doing so would create a dynamic and deeper educational experience for other users than pure annotation, which is commonly found in existing 3D visualization platforms such as SketchFab and Voyager.

%For instance, we provide interfaces that allow users to develop a custom shader to reveal certain surface details. Similarly, Resurrect3D allows users to integrate custom annotation, metering, and image processing algorithms into the workflow.

\paragraph{Customizability.} Resurrect3D also aims to provide a clean interface to allow domain experts to develop custom tools to visualize and analyze cultural artifacts. For instance, when a painting with pentimenti is captured through multispectral imaging, experts could implement a custom Principle Component Analysis tool to reveal certain surface details that are visible only with specific combinations of spectral bands, and then implement a custom shader to visualize different bands simultaneously in order to see the steps of the pentimenti's coming into being. In contrast, existing visualization platforms for cultural heritage generally do not provide the customizability, handicapping domain experts.

%We leverage the programming interface to develop a set of built-in tools such as metering, relighting, curtain view, which not only are readily useful on their own but also serve as examples for custom extensions by users.

\subsection{Deployment}

Resurrect3D is open source, allowing institutions to control the dissemination of digitized surrogates of their collection materials. Resurrect3D is live at: {\color{blue} \url{https://resurrect3d.lib.rochester.edu/}}, and the code is available at {\color{blue} \url{https://github.com/rochester-rcl/resurrect3d}}.

We have used Resurrect3D for a range of pedagogical, research, and entertainment applications. For instance, in the Lazarus project \cite{lazarus} we imaged and visualized, in Resurrect3D, the New York Public Library's Hunt-Lenox Globe~\cite{globe}, a minuscule (5 inches in diameter) globe depicting the Americas and one of the oldest terrestrial globes in existence. Custom relighting and image analyses in Resurrect3D reveal surface details not apparent on the globe. Resurrect3D is also used by the Ward project~\cite{ward} to visualize biological specimens from the Ward's Natural Science Establishment~\cite{wardsci}. Annotation and metering tools in Resurrect3D allows students and researchers to better understand the physiology of Ward's rare specimens.

This paper describes the system design of Resurrect3D. We focus on design decisions we made in making Resurrect3D accessible to casual users and domain experts, and highlight how the customizability can enable artifact-specific visualization and analysis.

\section{Related Work}

\paragraph{Generic 3D Visualization.} Arguably the most popular Web-based 3D visualization platform is SketchFab~\cite{sketchfab}, which is used by several cultural heritage institutions for sharing 3D models of their collections. SketchFab, however, is not without limitations. First, it is not open source, and requires a subscription-based payment for uploading and sharing works. Second, and more importantly, SketchFab is a generic visualization platform not designed with cultural heritage in mind and lacks field-specific functionality. In contrast, Resurrect3D provides a suite of tools that are more commonly used in cultural heritage. These include interactive measurement, interactive relighting material editing to reveal surface details, image analysis, and annotation tools.

\begin{figure*}[t]
\centering
\includegraphics[width=2\columnwidth]{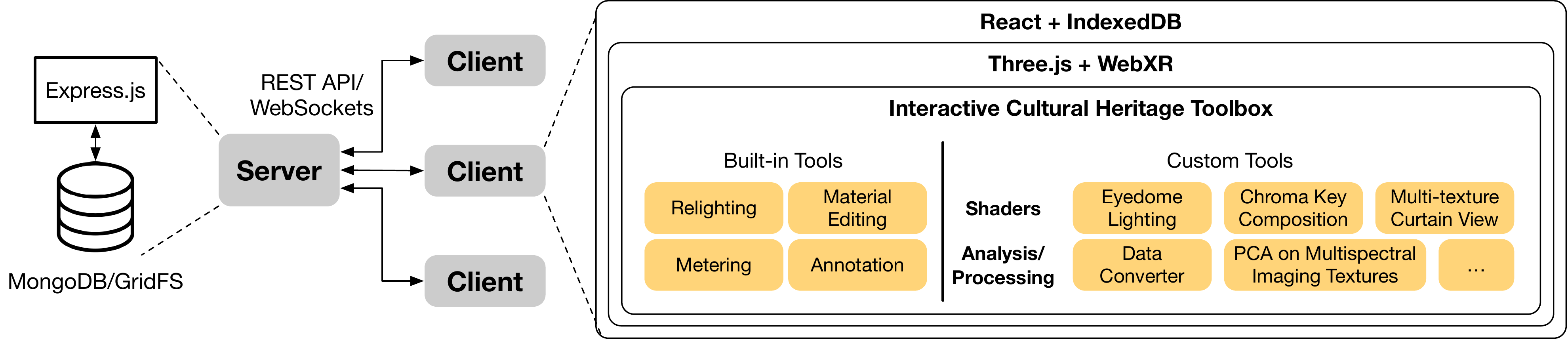}
\caption{Overview of the Resurrect3D system architecture.}
\label{fig:system}
\end{figure*}

\paragraph{Image Viewers for Cultural Heritage} There are a handful of image viewers designed for cultural heritage. They are usually highly specialized for working with Single Camera Multiple Light (SCML) data, which is obtained through Reflectance Transformation Imaging (RTI)~\cite{malzbender2001polynomial} and Portable Light Dome (PLD)~\cite{hameeuw2005easy} and is useful in revealing surface details. Notable desktop tools include RTIViewer~\cite{chi} and PLDviewer~\cite{pldviewer}. Web-based tools include Relight~\cite{ponchio2018compact}, Villanueva et al.~\cite{villanueva2019web}, the Oxford RTI Viewer~\cite{oxrti}, and  Pixel+~\cite{vanweddingen2020pixel+}. M.A.RL.I.E~\cite{jaspe2021web} supports not only RTI images but also shape and material representations with a flexible annotation system.

Resurrect3D supports SCML through Polynomial Texture Map (PTM)~\cite{malzbender2001polynomial}, which is a form of SCML data. The main difference between Resurrect3D and the SCML image viewers is that Resurrect3D supports other 3D modeling data and allows for developing custom tools to analyze 3D objects.
%, whereas SCML viewers provide little custom analysis and processing capabilities.
%are two-fold. First, Resurrect3D is fundamentally a 3D visualization platform that allows users to interactively explore an artifact whereas tools above are 2D image viewers. Second,

\paragraph{3D Visualization for Cultural Heritage.} Smithsonian has recently developed Voyager~\cite{voyager}, an open-source 3D Web-based visualization platform. Much like Resurrect3D, Voyager provides material editing, relighting, measurement, and annotation tools. CyArk provides a proprietary 3D viewer for their collections~\cite{cyark}. The viewer is designed primarily for viewing 3D models without many of the features that Resurrect3D provides such as relighting and material editing.

Resurrect3D differs from these two systems in two ways. First, Resurrect3D permits developing custom visualization (shading) and analysis tools, whereas neither system above has native support for customizability. We leverage the customizability and develop a set of tools such as Eyedome lighting and curtain view of multispectral textures, which are commonly used in cultural heritage visualization but are unavailable in either system. Second, both systems focus on visualizing mesh-based models whereas Resurrect3D supports a range of other data modalities commonly used in cultural heritage digitization such as multispectral imaging and SCML.

3DHOP~\cite{potenziani20153dhop} is a Web-based programming framework for 3D visualization, mostly used for cultural heritage. Through a declarative programming interface, 3DHOP provides a set of configurable knobs such as setting up camera and light parameters, picking points on the 3D model, and setting mesh transformation and visibility. These configurations are friendly to developers who are not familiar with graphics programming, but also limit the customizability to only what the configurable knobs offer. Resurrect3D, in contrast, provides customizability at a higher degree: we allow developers to directly write custom shaders and analysis tools. This provides more control over visualization and analysis, and thus allows for custom visualization such as multi-texture curtain view that is difficult to realize in 3DHOP.

\paragraph{Gamification} Serious games for cultural heritage deliver pedagogical goals~\cite{anderson2010developing, andreoli2017framework, bellotti2013serious, antoniou2013approach}. While Resurrect3D itself is not a game, the visualization, interaction, and customization capabilities of the platform provide opportunities to develop series games for real cultural heritage sites and artifacts, which is our future work.

\section{System Overview}

%This section first outlines the principles that guide the design of Resurrect3D, followed by a  description of the systems architecture, including the server and the client side.

%As a visualization platform, Resurrect3D strives to provide a smooth, straightforward user interface for casual use-cases (e.g., virtual tourism), which we see as a bare minimum. On top of that, we have two technical goals: portability for developers and customizability for domain experts.

The rendering system is built using the classic server-client architecture. \Fig{fig:system} shows an overview of the system.

\paragraph{Server.} The server application is built using Express.js and Node.js for interfacing with the client requests and uses MongoDB as the backend database. Small metadata, such as annotations and session-specific settings (e.g., camera pose), are directly stored in MongoDB, while large objects (e.g., raw point clouds and texture maps) are stored using GridFS, which builds on top of MongoDB and is optimized for accessing large objects. 

Clients communicate with the server in two ways, depending on the scenario. When a large chunk of data is requested, such as the initial request of a 3D model, the communication is through REST API, which is optimized for throughput. In contrast, we use WebSocket for subsequent communications that exchange small amounts of data but require real-time response, as WebSocket is optimized for latency. A typical scenario is when a teacher client sends her current state (e.g., camera pose, annotation) to student clients in the presentation mode.

\paragraph{Client.} The client-side UI is built using the React library~\cite{reactjs}. The rendering system is built on the three.js library~\cite{threejs} to leverage WebGL for GPU acceleration. The event-driven system in three.js integrates well with the asynchronous nature of React, allowing us (Resurrect3D developers) and future contributors to focus on developing functionalities specific to the cultural heritage domain. Finally, Resurrect3D also leverages WebXR to provide support for basic VR rendering of 3D objects.

We support two forms of client interaction. First, clients can perform completely independent interactions and analyses on an object. Second, clients can communicate with each other through the server, e.g., synchronizing different users to share the current state of one user's view. This can be useful in a classroom lecturing setting where students want to see what the teacher is seeing.

%Resurrect3D provides a set of built-in functionalities that allow users to interact with 3D models. These include basic object manipulation tools for panning, rotating, and zooming, as well as a capture tool for capturing a downloadable image of the object. Note that the capture is not simply a screenshot of the computer screen; rather, it captures an image in the virtual camera’s camera space, mimicking a photo that a user would have taken when viewing the object in real life. Resurrect3D also provides an export tool for downloading a 3D model as an STL or OBJ file along with its various metadata.

Finally,  we also implement a basic client-side caching system using the IndexedDB API, which stores compressed model data and drastically cuts down on the load times and bandwidth required compared to always fetching the model directly from the server.

\section{Interactive Cultural Heritage Toolbox}

%The first of the two is a composite of Principle Components  2, 8, and 22 on the full set of both reflective and fluorescent bands. The second is a Blur and Divide (blurred background divided by snap foreground composed of a band ration of a fluorescent image and an IR.

We first describe the design methodology that we adopt in developing Resurrect3D. We then describe Resurrect3D's cultural heritage toolbox which includes not only a set of built-in tools, but also generic programming interfaces that allow expert developers to develop custom analysis and visualization tools.

\subsection{Development Cycle and Design Methodology}

We worked closely with researchers spanning several cultural heritage disciplines including archaeology, classics, history, and natural science. In particular, our development cycle is the following:

\begin{enumerate}
	\item Users describe an ideal requirement to developers. In this step, users do not specify the exact implementation; rather, they focus on describing the expected utility of a feature.
	\item Given the feature request, developers quickly prototype a tool, and demonstrate the tool to the users. In this step, developers focus on delivering the required functionality without paying much attention to the UI design. Users provide feedbacks  until the functionality meets the requirement.
	\item The developers enter the next phase of the development cycle, where they focus on the UI design, aiming to provide an intuitive and responsive user interface of all technical skill levels without compromising the utility of the tool.
\end{enumerate}

For instance, a researcher in the University of Rochester's Department of Religion and Classics had expressed interest in a feature that allowed them to change the color of specific sections of an object's texture to improve readability of certain details. The team then prototyped a chroma key shader using the platform’s existing framework for custom tools and was able to share it with the researcher and acquire feedback in a short period of time. In fact, this is another reason we choose to make Resurrect3D a browser-based platform, as the team can iterate and acquire feedback much faster than with native applications. 

The UI is developed organically, taking many cues from established conventions in 3D graphics software for implementing the controls. As more custom tools add added, the team develops UI elements (sliders, color pickers, toggle buttons, etc) incrementally. A small group of faculty and student test users would then provide feedback on their user experience and the team would attempt to address any issues. The UI elements created through this process then form the building blocks of the platform's custom UI integration.

%Users uploading their own content can customize which tools are available to the end user, as the lighting, shading, metering, annotation, sharing, and exporting tools can all be disabled or enabled on a per model basis. A user can also persist all settings, from lighting positions to shader uniforms, for their uploaded models. We have developed a standard pipeline and design system for implementing such tools, focusing on extensibility and ease-of-use for prototyping new processing and analysis tools such as the experimental curtain view feature described below.

\begin{figure*}[t]
\centering
\includegraphics[width=2\columnwidth]{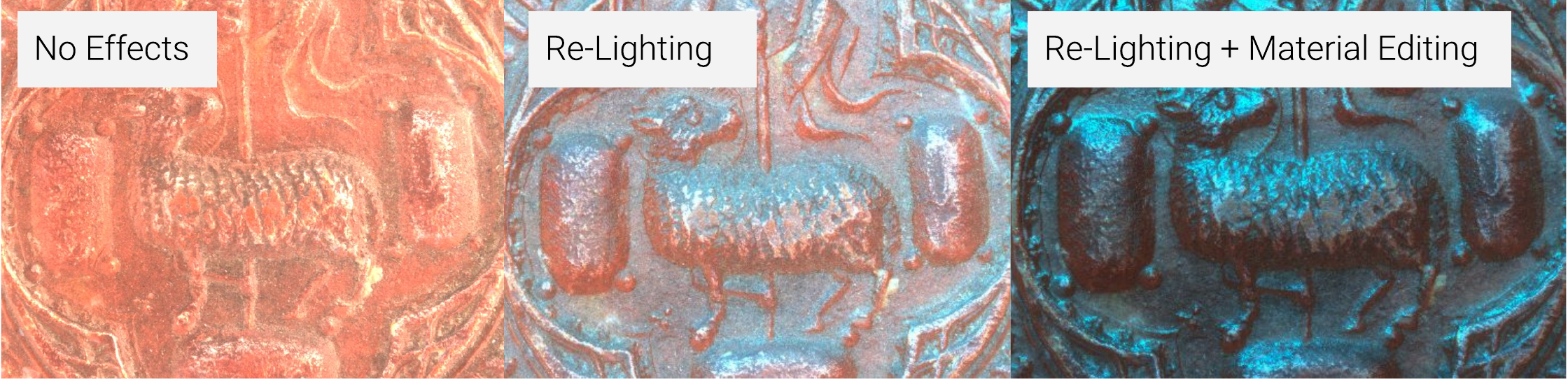}
\caption{A small wax seal from University of Rochester's Rare Books, Special Collections, and Preservation department. This example shows a combination of re-lighting and material editing (right) best increases the visibility of the seal's finer details compare to relighting alone (middle) and the oroginal rendering without any digital surface-revealing techniques (left).}
\label{fig:relighting}
\end{figure*}

\begin{figure}[t]
  \centering
  \includegraphics[width=.5\columnwidth]{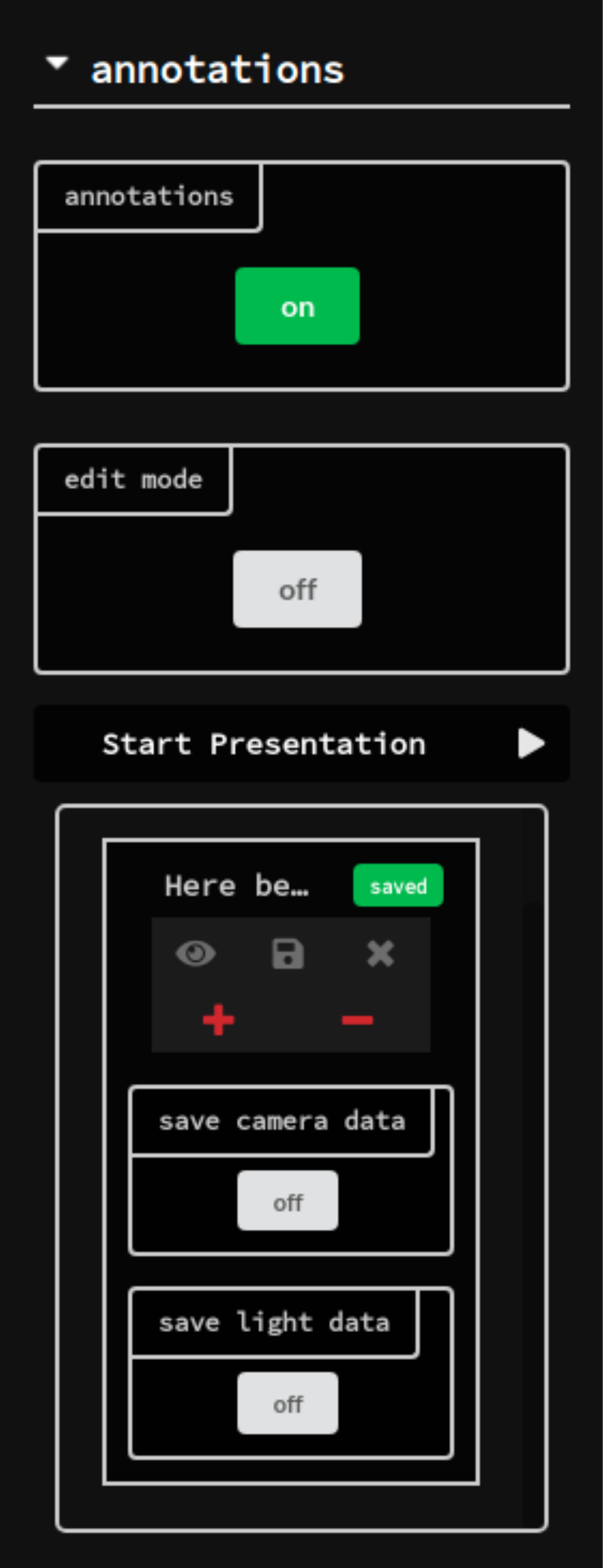}
  \caption{The UI for the presentation/story mode.}
\label{fig:story}
\end{figure}

\subsection{Basic Tools for Cultural Heritage}

\paragraph{Revealing Surface Details} A key requirement for cultural heritage visualization is to examine/reveal the surface details of an artifact to see the ``invisibles''. To that end, Resurrect3D provides a set of interactive tools including relighting to material editing.

A well-known technique used by historians and archaeologists is to point a light source to a specific area of the surface to bring out the details. Resurrect3D allows users to digitally emulate this process by specifying various properties of the light source (e.g., color, intensity) and to move the light source around the scene to relight specific regions.

%The prerequisite of relighting is precise surface normals. Resurrect3D supports three ways to obtain a normal map.
%\begin{itemize}
%    \setlength\itemsep{-1pt}
%	\item Directly import an existing normal map.
%	\item Infer the normal map from the texture map through Sobel filtering. This is useful when a direct normal map is unavailable.% Our actual implementation ports the algorithm implemented in the nvidia-texture-tools library~\cite{nvtexturetool}.
%	\item Generate a normal map from a  Polynomial Texture Mapping (PTM) file~\cite{malzbender2001polynomial}. This enables RTI for better relighting.
%\end{itemize}

Another useful method for revealing surface details is through editing the surface material of an artifact, such as the normal scale, metalness, and roughness values. For instance, changing the metalness essentially changes the albedo of the material, allowing fine-grained geometry to come out easily under certain light sources.

Resurrect3D allows users to arbitrarily combine interactive material editing and relighting to best bring out the details. \Fig{fig:relighting} compares how a proper combination of relighting and material provides the best visual details.

\paragraph{Annotation and Metering.} Cultural heritage experts enhance artifacts by adding metadata (e.g., texts, audio, images, video), which can later be disseminated for research or pedagogical purposes. Resurrect3D allows users to add annotations in a positional way by associating annotations with a particular position on the 3D model.

Positional annotation is implemented through a React callback that passes the click position to three.js, which translates the position to the coordinates in the camera space, which, in turn, get reprojected to the world space. This allows the annotation to be associated with the particular position of the model, invariant with the camera pose.

%The annotations are rendered with three.js’s CSS2DRenderer, which uses the CSS transform property to apply 2D translations to DOM elements based on camera position. The output of the CSS2DRenderer is overlaid on the WebGLRenderer’s canvas element.

Digitizing artifacts brings an inherent benefit: accurate metering of the geometric properties of an artifact, such as the distance between two points, the surface area, or the volume. This metering capability is important especially for objects of extreme scales (large or small) or brittle under any physical measurement.

%Resurrect3D provides an interactive measurement tool with multiple units. This is achieved using the same underlying implementation of position annotation: translating any selection from the user to coordinates in the camera space, which are then translated to the world space of the model, in which the actual measurement calculation, be it distance or area, is carried out.

\paragraph{Interactive Storytelling.} Resurrect3D facilitates cultural heritage education by allowing one user to create a story, i.e., a saved session of annotations, measurements, material and lighting settings. The story can later be launched by another user to experience a virtual walkthrough of the 3D model with guided tools. At any point during the walkthrough, the user can apply custom interactions with the model, essentially exploring the virtual space on their own. We find this useful for students to validate/critique the pre-recorded instructions.

\Fig{fig:story} shows the UI component for the presentation mode. A user can reorder the annotations in a sequence while choosing which data to persist alongside the annotation (e.g., the camera or lighting position, shader settings). Other users can later cycle through the annotations. The main camera interpolates between the positions stored alongside the annotations, creating an animated ``guided tour'' of the object.
%The team is also working on implementing the storytelling feature in VR, allowing for a more immersive experience for the user.

%The interactive storytelling feature will be expanded upon in the new ``lecture mode'', where a user can stream their navigation of an object in real-time. Camera position and other settings are serialized and sent to subscribers' viewer instances using WebSockets, allowing for dynamic updates of their interface.

\subsection{Customizability}

A key differentiating feature of Resurrect3D is that it provides great flexibility for expert users to design custom tools. The interfaces for custom tooling can be classified into two main categories: 1) custom shaders, which allow visualizations of the 3D model different from the default shaders in three.js, and 2) custom analysis tools, which allow for statistical, image, and vision processing on the model/metadata. Enabling custom analysis tools extends Resurrect3D from a visualization platform to also be an analytics platform.

\subsubsection{Custom Shaders}

The cultural heritage community, over the years, has designed creative shading techniques to aid in viewing artifacts. Resurrect3D provides a straightforward interface for developing custom shaders. This is accomplished by leveraging three.js’ capability of directly integrating a shader written in OpenGL Shading Language (GLSL).

The design goal of the interface is to allow developing the shaders without worrying about how the shader code will be integrated into the React application. To that end, we provide a template, which defines a ``loader'' function that takes a three.js object and adds the custom shader to the three.js namespace. The code snippet below illustrates this template. The only code that the programmers have to provide is the shader itself, which is a plain JavaScript object containing the uniforms and the GLSL shaders, which graphics programmers should be familiar with.

\begin{lstlisting}[language=JavaScript,numbers=left,aboveskip=\smallskipamount,belowskip=\smallskipamount]
export default function loadNewShader(threeInstance: Object): Promise {
  return new Promise((resolve, reject) => {
    threeInstance.NewShader = {
      //user-define custom shader below
      uniforms: {...},
      vertexShader: ...,
      fragmentShader: ...
    }
    resolve(threeInstance);
  });
}
\end{lstlisting}

To improve performance, the loader function integrates the GLSL shader code with the main React UI through JavaScript Promises. That is, the loader function returns a Promise that resolves the three.js object. We use Promises for two reasons. First, it exposes an interface that is asynchronous by nature, which allows for additional functionality such as loading custom shaders from external sources (i.e. from a separate server or a content distribution network). Second, loaders can also be chained together with other asynchronous functions in the main React component (e.g., loading mesh data from the server) or executed in parallel (using the Promise.all method).

\paragraph{Automatic UI Integration.} A custom shader and its various knobs, once defined, have to be added to the main application UI so that users can decide when and how to apply the shader. To allow developers to focus on the shader design without having to worry about the UI design, Resurrect3D exposes two APIs for programmers to specify the knobs for each custom shader; Resurrect3D then automatically adds the knobs to the application UI.

The first API allows programmers to define a UI group for a shader; all the knobs will show up in the group.

\begin{lstlisting}[language=JavaScript,numbers=left,aboveskip=\smallskipamount,belowskip=\smallskipamount]
shaderGroup = new ThreeGUIGroup("shaders");
\end{lstlisting}

The second API allows programmers to specify a UI component with its name, type, and the various properties of the component, of which the main property is the callback functions to be called when corresponding user interactions are triggered.

\begin{lstlisting}[language=JavaScript,numbers=left,aboveskip=\smallskipamount,belowskip=\smallskipamount]
shaderGroup.addComponent("intensity", components.THREE_RANGE_SLIDER, {
  //define the callback when a user slides the bar
  callback: ...,
  // define properties of a slider
  min: 0.0,
  max: 4.0,
  step: 0.1,
});
\end{lstlisting}

We provide three custom shaders in the current design of Resurrect3D. These shaders are readily useful on their own, but also serve as examples for user extensions.

\begin{figure}[t]
  \centering
  \includegraphics[trim=0 0 0 0, clip, width=\columnwidth]{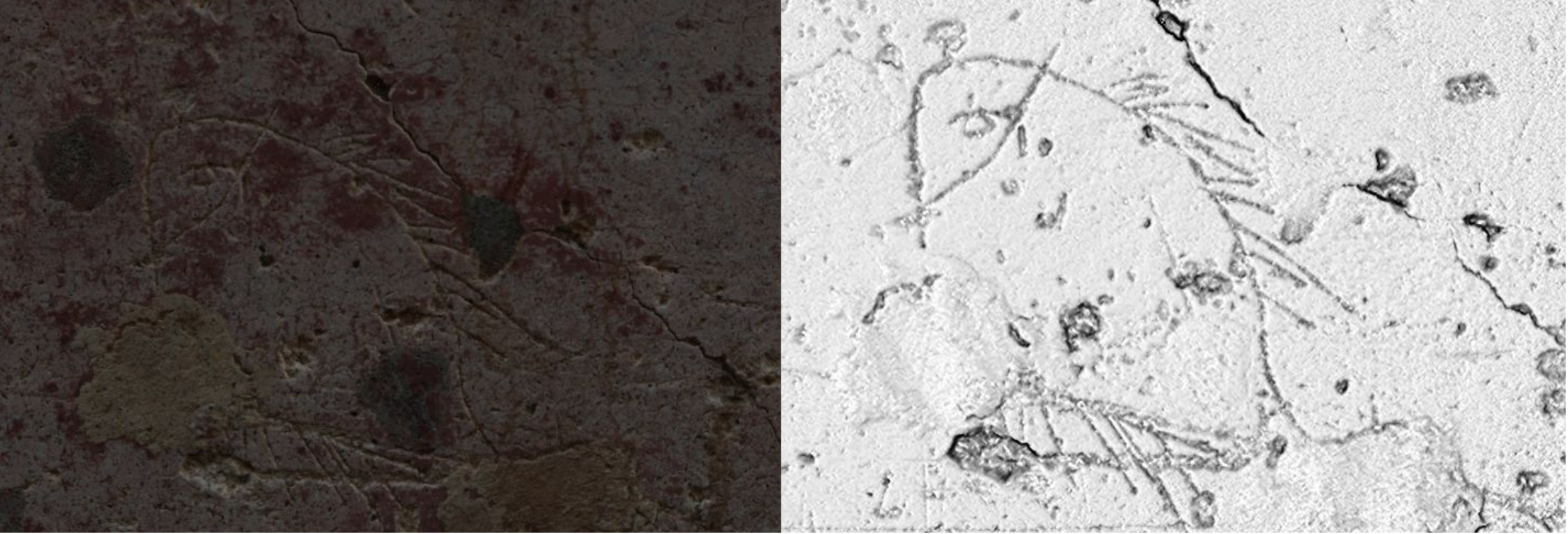}
  \caption{A Graffito in Oplontis Villa A in Torre Annunziata, Italy. The fish on the graffito (left) is much more visible with the EDL shader (right).}
  \label{fig:edl}
\end{figure}

\paragraph{EDL.} Relighting is computationally intensive if the object is to be rendered photo-realistically. In many use-cases, however, photorealism is less important as long as surface details are distinguishable. A typical technique to achieve that is Eye-Dome Lighting (EDL), which is an image-space technique to assign different colors to pixels depending on the depth information. EDL is commonly used for quick detail revealing~\cite{deibe2019supporting, hofer2018end}. Resurrect3D implements an interactive EDL shader following the algorithm by Boucheny~\cite{boucheny2009visualisation}. \Fig{fig:edl} shows the effect of our EDL shader in bring out the details of the fish on a Graffito in Oplontis, Italy.
%, some implementation details adapted from both the Potree and Cloud Compare projects.

\begin{figure}[t]
  \centering
  \includegraphics[trim=0 0 0 0, clip, width=\columnwidth]{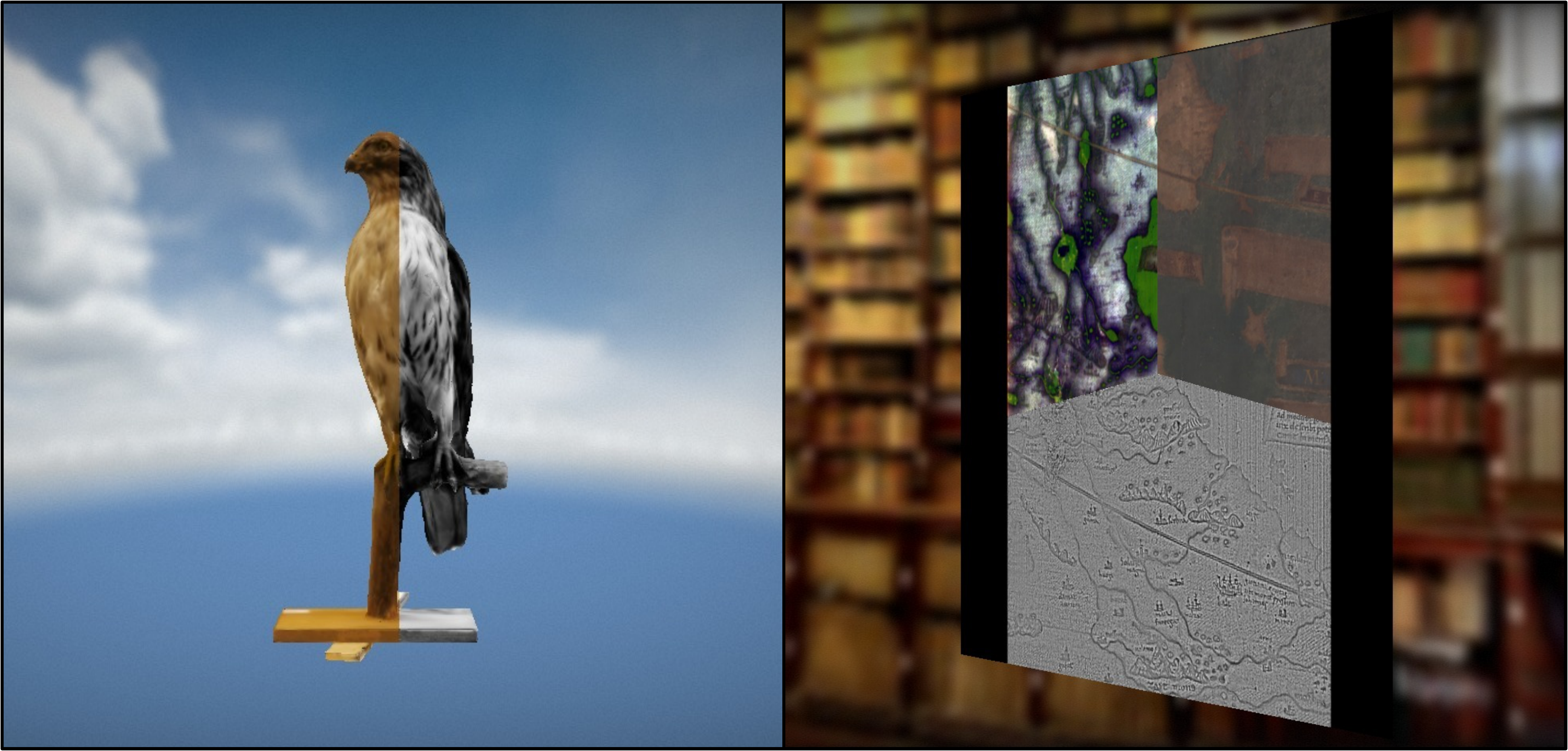}
  \caption{Curtain-view mode. A redtail hawk from Ward's collection visualized with two texture maps (left). World map by Henricus Martellus with three color bands from multispectral imaging (right).}
  \label{fig:curtain}
\end{figure}

%\begin{figure}[t]
%\centering
%\includegraphics[width=.5\columnwidth]{edl}
%\caption{A Graffito in Oplontis Villa A in Torre Annunziata, Italy. The fish on the graffito (left) is much more visible with the EDL shader (right).}
%\vspace{-10pt}
%\label{fig:edl}
%\end{figure}
%(courtesy of the River Campus Libraries' Digital Scholarship department)

\paragraph{Chroma Keying.} To improve the contrast and readability of an object's surface details, historians often use an idea from the visual effects industry called Chrome Keying, which replaces a specific color with another color. Resurrect3D provides a custom Chroma Key shader, which allows users to change colors of arbitrary points on the mesh and to adjust the ratio between the original color and the replacement color for finer control over the end result.

%pick a specific color on the mesh and replace it with another color of their choosing, improving the contrast and readability of the object’s surface details. The user can also 

\paragraph{Multi-texture Curtain View.} Simultaneously examining multiple bands from multispectral imaging is a common strategy to explore different elements of the object. For instance, many paintings have pentimento, a first sketch that masters make on the canvas and that later is painted over and changed; multispectral imaging captures different layers on the painting as different images, examining which allows people to see backward in time what an object once was and to view the steps of its coming into being.
%It was one of these pentimenti that, in 2005, revealed the painting known as Salvator Mundi to be by Leonardo Da Vinci.

%\begin{figure}[t]
%\centering
%\includegraphics[width=.5\columnwidth]{curtain}
%\caption{Curtain-view mode. A redtail hawk from Ward's collection visualized with two texture maps (left). World map by Henricus Martellus with three color bands from multispectral imaging (right).}
%\vspace{-10pt}
%\label{fig:curtain}
%\end{figure}

Through a custom shader, Resurrect3D provides what we call the curtain-view mode, which samples different regions of the rendered frame from different textures. As the mouse or touch gesture moves, the boundaries of the regions move accordingly, allowing users to understand and explore the history behind an artifact. \Fig{fig:curtain} shows two artifacts in the curtain-view mode: a redtail hawk model from the Ward project~\cite{ward} and the famous world map by Henricus Martellus with three color bands from multispectral imaging~\cite{van2018henricus}.
%This shader can also be easily extended to visualize different processed versions of the object. 

\subsubsection{Custom Processing and Analysis Tools}

Oftentimes experts analyze and process the digital artifact before visualization. For instance, one could perform a Principle Component Analysis (PCA) on different bands captured through multispectral imaging to generate new texture maps for visualization.

Resurrect3D leaves a generic interface for developing custom analysis tools by exposing the model and metadata of an artifact to developers. The analysis tool and the main rendering code in three.js live in the same namespace. Thus the analysis tools can easily interoperate with the rendering pipeline.

Much like the custom shaders, the analysis tools also expose a simple Promise-based interface. The code below shows the programming interface, where the analysis function takes either a Mesh and/or its metadata, so that the function has access to all underlying geometry and material data, and returns a Promise that resolves an object with the same type as the input:

\begin{lstlisting}[language=JavaScript,numbers=left,aboveskip=\smallskipamount,belowskip=\smallskipamount]
function myAnalysisFunction(object: THREE.Mesh | THREE.Group): Promise {
  return new Promise((resolve, reject) => {
    // define the analysis function here
    resolve(object);
  });
}
\end{lstlisting}

% Since the function returns a Promise, it is asynchronous in nature, so any processing can be offloaded to a WebWorker, or even a server-side application.

As an example, we provide perhaps the simplest example of an analysis tool, which imports and converts common 3D formats into a format that is required by the visualization system. Specifically, since Resurrect3D by default supports a physically based rendering (PBR) workflow, the conversion tool converts 3D formats (e.g., OBJ, FBX, VRML) into a range of PBR maps (e.g., diffuse and displacement maps) in the JSON Object Scene Format required by three.js.  Wherever possible, we make use of the WebWorker~\cite{webworker} to parallelize and offload processing to background threads.
%The PBR maps will also be compressed and sent back to the server for other clients to consume. 

\section{Conclusion}

More than ninety percent of the objects in museums around the world will never be seen by the public, either for the lack of display space or because of damage or fragility. We envision a future where visualization platforms turn every museum and archive into an open university. To achieve that vision, Resurrect3D provides not only the basic visualization and interaction capabilities, but also the customizability that allow domain experts to develop artifact-specific analysis and visualization tools.

\bibliographystyle{ACM-Reference-Format}
\bibliography{refs}

\end{document}